\documentclass{article}
\usepackage{spconf,amsmath,epsfig}
\usepackage{float}
\usepackage{caption}
\usepackage{algorithm}
\usepackage{algpseudocode}
\usepackage{booktabs}
\usepackage[numbers,sort&compress]{natbib}
\usepackage{amsfonts,amssymb}
\usepackage{graphicx} 
\usepackage{epstopdf}
\usepackage{verbatim}
\usepackage{graphicx,times}
\usepackage{subfigure}
\usepackage{natbib}
\usepackage{amssymb,amsmath}
\usepackage{bm}
\usepackage{color}

\usepackage{amsmath} 

\title{Weighted Hierarchical Sparse Representation for Hyperspectral Target Detection}
\name{Chenlu Wei, Zhiyu Jiang\sthanks{2020 IEEE. Personal use of this material is permitted. Permission from IEEE must be obtained for all other uses, in any current or future media, including reprinting/republishing this material for advertising or promotional purposes, creating new collective works, for resale or redistribution to servers or lists, or reuse of any copyrighted component of this work in other works. $*$Corresponding author: Zhiyu Jiang (jiangzhiyu@nwpu.edu.cn).}, Yuan Yuan}
\address{\small School of Computer Science and Center for OPTical IMagery Analysis and Learning (OPTIMAL), \\
\small Northwestern Polytechnical University, Xi'an 710072, Shaanxi, P. R. China.\\
        }

\begin{document}
%

\maketitle
\begin{abstract}
Hyperspectral target detection has been widely studied in the field of remote sensing.  However, background dictionary building issue and the correlation analysis of target and background dictionary issue have not been well studied. To tackle these issues, a \emph{Weighted Hierarchical Sparse Representation} for hyperspectral target detection is proposed. The main contributions of this work are listed as follows.
1) Considering the insufficient representation of the traditional background dictionary building by dual concentric window structure, a hierarchical background dictionary is built  considering the local and global spectral information simultaneously.
2) To reduce the impureness impact of background dictionary, target scores from target dictionary and background dictionary are weighted considered according to the dictionary quality. 
Three hyperspectral target detection data sets are utilized to verify the effectiveness of the proposed method. 
And the experimental results show a better performance when compared with the state-of-the-arts.

\end{abstract}



\section{Introduction}
\label{sec:intro}

Hyperspectral images (HSIs) contain more abundant spectral information than natural images,
which is a unique advantage for target detection in HSIs
\cite{Du2019, Du2019_Autoencoder}. 
Hyperspectral target detection is aimed at locating target pixels in HSIs given the target spectral signature from the spectral library. In other words, it is a binary classification problem and each pixel is labelled as the target or the background. Hyperspectral target detection can be applied widely in military and civilian domain, such as military target detection, environmental monitoring, rock and mineral identification, etc \cite{HSDD}.

In the past decades, numerous methods have been proposed for hyperspectral target detection.
Quite a few linear mixing model (LMM)-based methods achieve good performance, such as CEM\cite{CEM}, MSD\cite{MSD}, ACE\cite{ACE}.
These methods usually get the output score by signalizing the target signature while restraining the background signature 
and based on the assumption that the noise obeys a zero mean multivariate normal distribution when estimating background \cite{Du2019}.
However, the assumption doesn't hold in real HSIs and is difficult to achieve an satisfactory detection performance.

Recently, sparse representation based methods have been widely used and achieved a satisfactory performance.
The sparse-based methods don't need to suppose the distribution of pixels, and it can alleviate the change of the target spectrum with different lighting and atmospheric conditions. 
Chen \emph{et al}. \cite{STD} proposed a sparse representation-based target detector (STD) and this is the first time that the sparse-based method was introduced for hyperspectral target detection.
And then, other methods based on sparse representations have sprung up,
such as Sparse Representation Based Binary Hypothesis (SRBBH) \cite{SRBBH}, Combined Sparse and Collaborative Representation (CSCR) \cite{CSCR}, and Hybrid Sparsity And Statistics-Based Detector (HSSD) \cite{HSSD}.
Chang \emph{et al}. \cite{Du2019_Autoencoder} introduced the sparse autoencoder to anomaly detection in HSIs based on deep learning.

However, there are still some issues with the method based on sparse representation, and solutions to these issues are proposed. 
1) In order to focus on the adaptive ability of the background dictionary, the traditional sparse representation based methods usually use a sliding double window to select the background dictionary. Whereas, those methods discard the rich and comprehensive global information of image.
To settle this issue, the pixels with low target scores are used to learn the global background dictionary by dictionary learning. Background dictionary based on global and local information can  reconstruct the pixels well. 
2) Because of the unbalanced quality of the target and background dictionary, it is not reasonable to directly use the error of reconstruction residuals to represent the target score. Weighting scores is an effective way to deal with unequal dictionary quality.
It is worth noting that because of the limitations of the available data sets, the current deep learning-based methods are mainly based on sparse autoencoder which still have the same issues as traditional methods. 

The rest of this paper is organized as follows. In Sec. 2, a brief presentation of the proposed method is expounded. In Sec. 3, the effectiveness of the proposed method is demonstrated. Finally, the conclusions are drawn in Sec. 4.

\begin{figure*}[htp]
	\centering
	\begin{minipage}[b]{1.0\linewidth}
		\centering
		\graphicspath{{picture/}}
		\centerline{\epsfig{figure=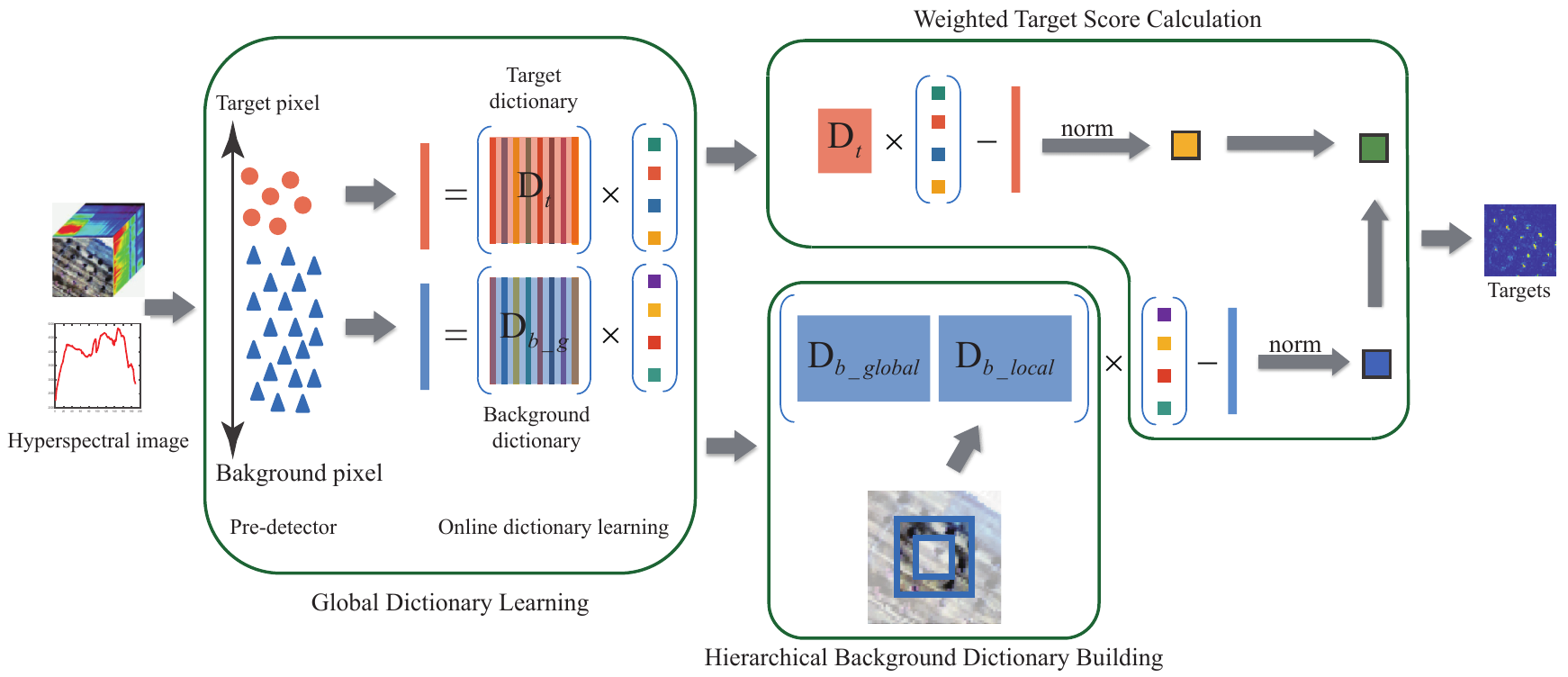,width=17cm}}
	\end{minipage}
	\caption{The flow chart of the proposed W-SHR. Firstly, the pre-detector and online dictionary learning methods are utilized to get the original global target and background dictionary.
	And then, a hierarchical background dictionary building method is proposed by considering the global and local background dictionary simultaneously.
    Finally, the target scores are obtained by specifying different weights to the scores from target and background dictionary.
	}
	\label{fig:flowchart}
	\vspace{-0.4cm}
\end{figure*}

\section{METHODOLOGY}
A method called \emph{Weighted Sparse Hierarchical Representation} (W-SHR) is proposed in this paper. In this section, the pre-detector and dictionary learning methods are introduced firstly. And then, a method to get a hierarchical dictionary is described. Finally, A reasonable way to get the target score of each pixel is expounded.
 
 
\noindent
\subsection{Global Dictionary Learning}

 The traditional method CEM \cite{CEM} is used to pre-detect in order to roughly know the score of each pixel, which is the basis for selecting the training sets of target and background dictionary. 
 The pixels with high scores are selected as the target training samples, and the low score pixels as the background training samples. 
Online Dictionary Learning (ODL) is an effective method to learn a dictionary representing one signal $\bm{x}$. It can minimize the local approximation of the expected cost and doesn't require the explicit learning rate tuning \cite{ODL}. In this paper, ODL is applied to learn a global target dictionary and a global background dictionary from the training sets selected by CEM.
Considering the single target and the messy background, the number of atoms in global background dictionary is much larger than that in the target dictionary.

\subsection{Hierarchical Background Dictionary Building}
 \vspace{-0.1cm}
In traditional sparse target detection algorithms, scholars always adopt sliding dual concentric window to obtain adaptive local background dictionaries.
Those methods fail to make full use of the entire HSIs because they ignore the overall attributes of the HSIs.
The method to get a hierarchical background dictionary $\bm{D}_b$ by concatenating the global background dictionary $\bm{D}_{b\_glo b a l}$ and the local background dictionary $\bm{D}_{b\_ l o c a l}$ is like Eq. (\ref{Eq:1}).
The background dictionary obtained has the characteristics of extensiveness and adaptability.
\begin{equation}
\label{Eq:1}
\bm{D}_b=\left[\bm{D}_{b\_g l o b a l}, \bm{D}_{b\_ l o c a l}\right].
\end{equation}

It is worth noting that the local dictionary should adhere to the constrain in Eq. (\ref{Eq:2}) before concatenation, which is consistent with the global dictionary by ODL.
\begin{equation}
\label{Eq:2}
\bm{D}_{b_{-}local} \in \bm{R}^{m \times k}, \text { s.t. } \forall j=1, \ldots, k, \bm{d}_{j}^{T} \bm{d}_{j}=1.
\end{equation}

\subsection{Weighted Target Score Calculation}
 \vspace{-0.1cm}
Target score is defined as the probability of being a target.
The target score of traditional sparse-based methods is error of reconstruction residuals.
However, the quality difference between background dictionary and target dictionary is usually ignored.
A novel method is proposed to resolve this issue in this section.


Sparse representation for hyperspectral target detection assumes usually that the background/target pixel can be reconstructed primely by background/target dictionary. The residuals of recovery a pixel $\bm{x}$ by the target dictionary $\bm{D}_{t}$ and the background dictionary $\bm{D}_{b}$ can be obtained according to Eq. (\ref{Eq:3}).

\begin{equation}
\label{Eq:3}
\begin{array}{l}{
 {\bm{r}_{t}(\bm{x})}=\left\|\bm{x}-\bm{D}_{t} \bm{\hat{\alpha}}_{t}\right\|_{2}} , \vspace{0.5ex}\\
 {\bm{r}_{b}(\bm{x})=\left\|\bm{x}-\bm{D}_{b} \bm{\hat{\alpha}}_{b}\right\|_{2}},

 \end{array}
\end{equation}

\noindent
where $\bm{\hat{\alpha}}_{t}$ and $\bm{\hat{\alpha}}_{b}$ are the recovered sparse vectors with L1 loss. The true target pixels can be recovered well with target dictionary while be recovered poorly with background dictionary.
The target score is calculated by the residuals from target dictionary and the residuals from background dictionary respectively like Eq. (\ref{Eq:4}).
\begin{equation}
\label{Eq:4}
\begin{array}{l}{
\bm{\text{S}}_{t}=\frac{\bm{r}_{t}(\bm{x})-\min \bm{r}_{t}(\bm{x})}{\max \bm{r}_{t}(\bm{x})-\min \bm{r}_{t}(\bm{x})}}, \vspace{1ex}\\
\bm {\text{S}}_{b}=\frac{\max \left(\bm{r}_{b}(\bm{x})\right)-\bm{r}_{b}(\bm{x})}{\max \left(\bm{r}_{b}(\bm{x})\right)-\min\left(\bm{r}_{b}(\bm{x})\right)}.\end{array}
\end{equation}

Different weights should be given according to the quality of dictionary, and the final target score is shown by  Eq. (\ref{Eq:5}).
\begin{equation}
\label{Eq:5}
{\text{S}=(1-\gamma)\bm{\text{S}}_{t}+\gamma\bm{\text{S}}_{b}},
\end{equation}

\noindent
where $\gamma$ is a constant between 0 and 1, which indicates the quality of the background dictionary against the target dictionary.
$\gamma$ can reduce the negative impact of impureness and weak reconstruction ability of background dictionary.

\section{EXPERIMENT and Analysis}
\label{sec:format}
\begin{figure}[htb]
\begin{minipage}[b]{1.0\linewidth}
  \centering
  \graphicspath{{}}
  \centerline{\epsfig{figure=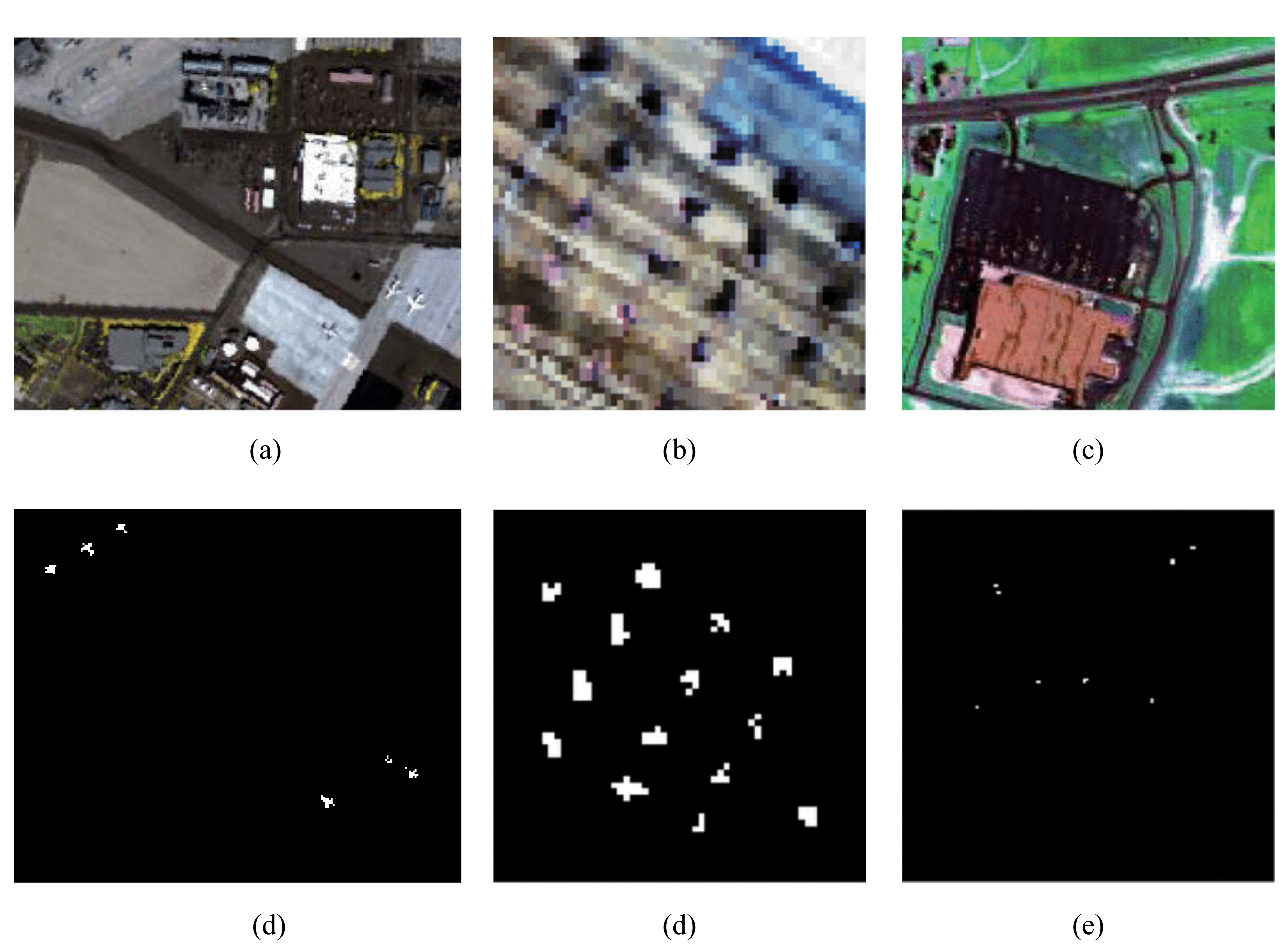,width=8.5cm}}
\end{minipage}
\vspace{-0.5cm}
\caption{Hyperspectral target detection data sets, including AVIRIS \uppercase\expandafter{\romannumeral1}, AVIRIS \uppercase\expandafter{\romannumeral2} and HYDICE. The false-color maps are showed in the first row and the labeled targets are showed in the second row respectively.}
\label{fig:datasets}
\vspace{-0.2cm}
\end{figure}

The proposed W-SHR is compared with the state-of-the-art methods in this section. 
The generic hyperspectral target detection data sets used were collected by AVIRIS sensor and HYDICE sensor as shown in Fig. \ref{fig:datasets}.
\cite{BCCR_2019}. Taking into account low SNR, water absorption, and bad bands in raw data, the bands mentioned above are removed just like others for the sake of fairness. 
The sizes of AVIRIS \uppercase\expandafter{\romannumeral1}, AVIRIS \uppercase\expandafter{\romannumeral2} and HYDICE are 240$\times$200$\times$189, 60$\times$60$\times$189, 150$\times$150$\times$162, respectively. 
The three numbers of the above sizes represent the width, height, and bands of data sets respectively.

The proposed W-SHR method is compared with the following state-of-the-art methods, CEM\cite{CEM}, ACE\cite{ACE}, STD\cite{STD}, CSCR\cite{CSCR}, and HSDD\cite{HSDD}. Moreover, for analyzing the contributions of this work, \emph{Sparse Hierarchical Representation}(SHR) is proposed as ablation experiment without weighting the scores. The evaluation criteria are Receiver Operating Characteristic (ROC) and Area Under Curves (AUC). 
The ROC curve in the upper left corner of the image indicates better performance.
Nevertheless, it is not easy to use ROC curves to visually compare the performance of methods when curves are close, scholars usually calculate the AUC to compare the performance of the various methods clearly.

For the three real hyperspectral data sets, 
the regularization parameter $\lambda$ and the sparsity $k$ are set to 0.1 and 5 respectively. The numbers of target and background training samples are set to 10 and 0.8 $\times$ N respectively, N is the number of all pixels in the data set. The numbers of global target and background dictionary atoms are set to 10 and 1000 respectively. According to experience, for AVIRIS \uppercase\expandafter{\romannumeral1}, the sizes of the OWR and IWR are 19 $\times$ 19 and 9 $\times$ 9 \cite{HSDD}, the weighting factor of background score $\gamma$ is set to 0.3. For AVIRIS \uppercase\expandafter{\romannumeral2},  the sizes of the OWR and IWR are 17 $\times$ 17 and 7 $\times$ 7 \cite{HSDD}, $\gamma$ is set to 0.2. For HYDICE data set, the sizes of the OWR and IWR are 15 $\times$ 15 and 5 $\times$ 5 \cite{HSDD}, $\gamma$ is set to 0.3.

\begin{figure*}[htb]
\begin{minipage}[b]{1.0\linewidth}
\centering
\graphicspath{{}}
\centerline{
    \subfigure[AVIRIS \uppercase\expandafter{\romannumeral1}]{\epsfig{figure=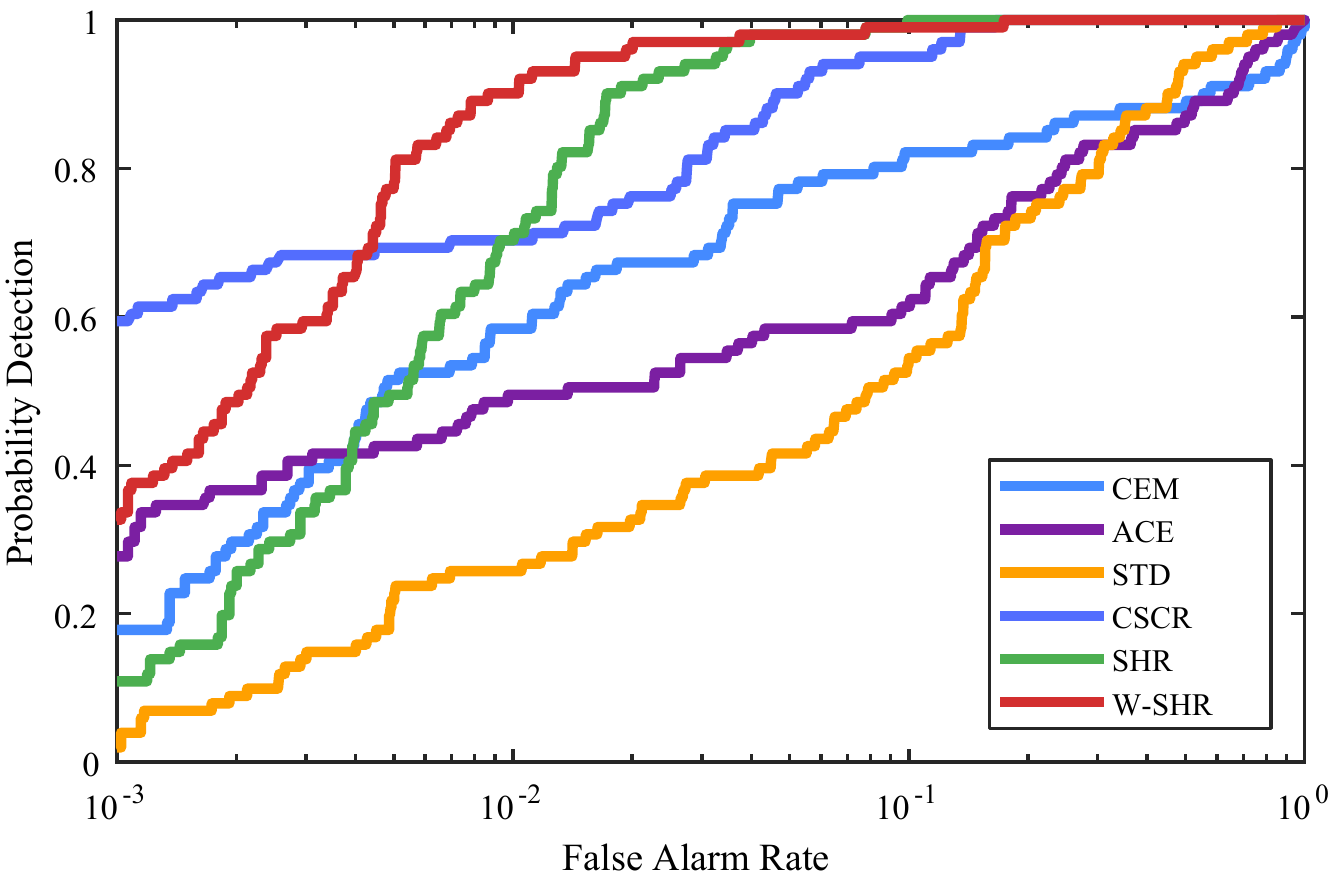,height=4.5cm,width=5.5cm}}
     \subfigure[AVIRIS \uppercase\expandafter{\romannumeral2}]{\epsfig{figure=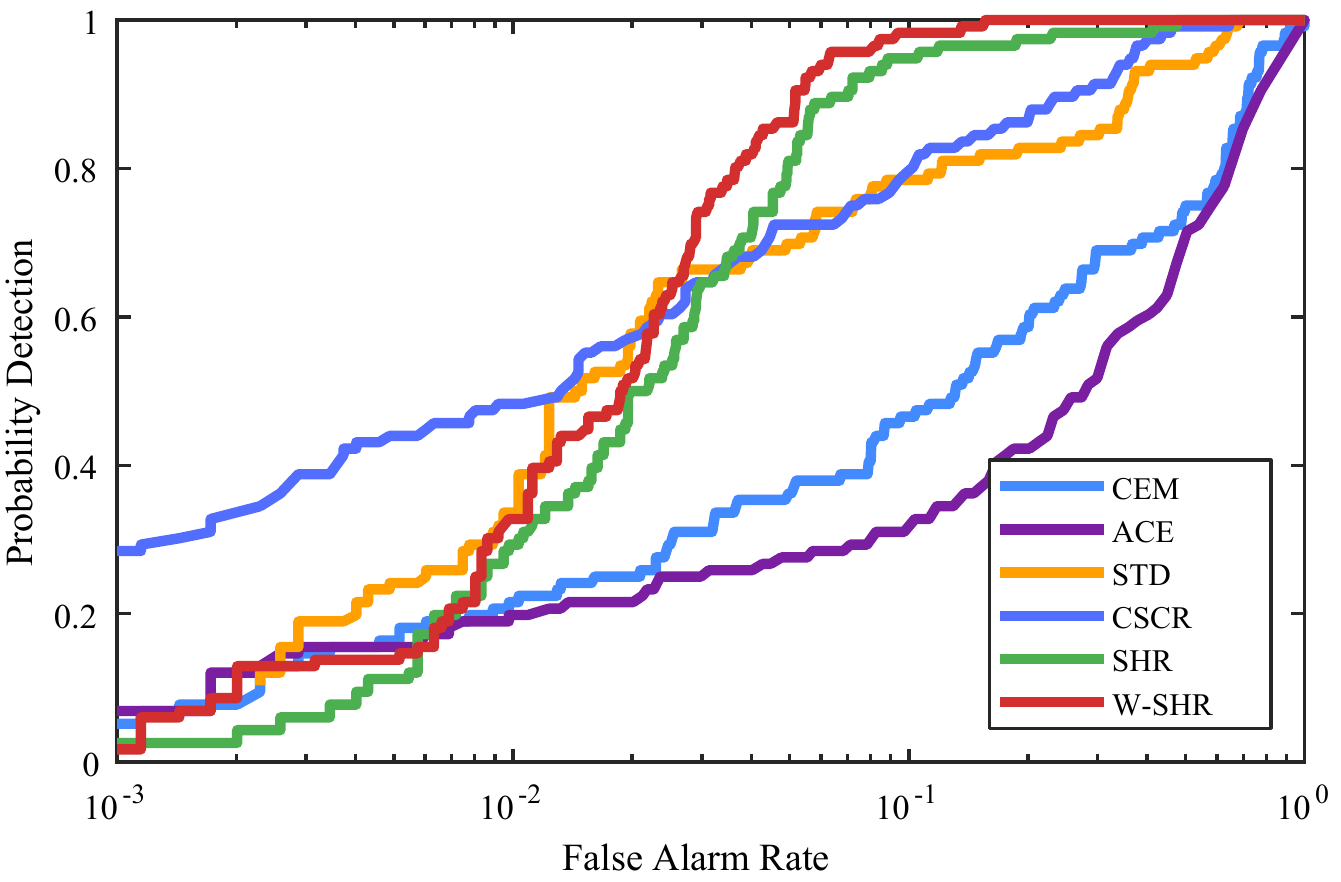,height=4.5cm,width=5.5cm}}
     \subfigure[HYDICE]{\epsfig{figure=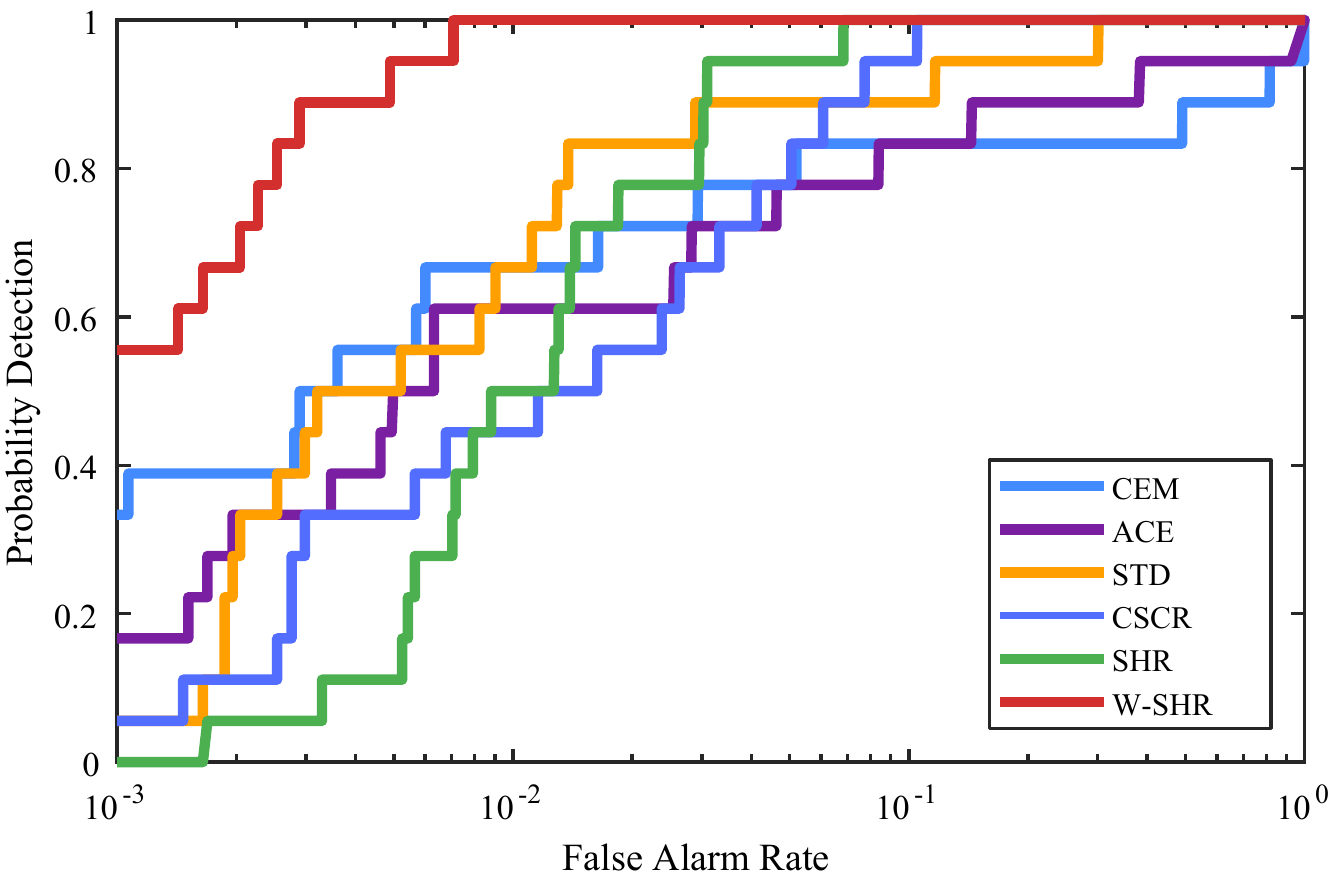,height=4.5cm,width=5.5cm}}
 }
\end{minipage}
\vspace{-0.4cm}
\caption{ROC curve results of the proposed method and competitive hyperspectral target detection methods on (a) AVIRIS I, (b) AVIRIS II, and (c) HYDICE.}
\label{ROC}
\vspace{-0.4cm}
\end{figure*}

The ROC curves on the three data sets are shown in Fig. \ref{ROC} and the AUC values are shown in Table \ref{AUC}.  For AVIRIS \uppercase\expandafter{\romannumeral1}, Although the ROC of CSCR has a higher detection rate when the false alarm rate is low, the ROC of W-SHR soon exceeds the ROC of CSCR and keeps probability detection consistently high. 
For a clearer comparison, the AUC of W-SHR surpasses other methods, which means the proposed method can perform well through combining the global and local background dictionary, and it can perform better through setting the reasonable weights.

\begin{table}[!hbp]
	\centering
	\caption{AUC values of the proposed method and competitive hyperspectral target detection methods on AVIRIS \uppercase\expandafter{\romannumeral1}, AVIRIS \uppercase\expandafter{\romannumeral2}, and HYDICE. Bolds represent the top two.}
	\vspace{-0.2cm}
	\begin{tabular}{cccc}
		\hline
		Methods & AVIRIS \uppercase\expandafter{\romannumeral1} & AVIRIS \uppercase\expandafter{\romannumeral2} & HYDICE\\
		\hline
		CEM\cite{CEM} & 0.8805 & 0.7389 & 0.8653\\
		ACE\cite{ACE}& 0.8453 & 0.6677 & 0.9043\\
		STD\cite{STD} & 0.8457 & 0.9090 & 0.9708\\
		CSCR\cite{CSCR} & 0.9843 & 0.9315 & 0.9738\\
		HSDD\cite{HSDD} & 0.9822 & 0.9590 & \textbf{0.9951}\\
		SHR & \textbf{0.9906} & \textbf{0.9625} & 0.9842\\
		W-SHR & \textbf{0.9940} & \textbf{0.9756} & \textbf{0.9984}\\
		\hline
	\end{tabular}
	\label{AUC}
\end{table}

For AVIRIS \uppercase\expandafter{\romannumeral2}, the local background dictionary inevitably contains many target pixels because there are multiple plane targets closely distributed on the entire image. Therefore, the weighting factor $\lambda$  is set smaller because of the quality reduction of the background dictionary. As shown in Table \ref{AUC}, W-SHR obtained the best results, exceeding HSDD, which is proposed to purify the background dictionary. It can be seen that W-SHR is an effective way to reduce the negative impact of background dictionary impurity.

For HYDICE, Because HSDD performed well on HYDICE and the AUC of HSDD reached as high as 0.9951 \cite{HSDD}, SHR failed to exceed it, but the weighted SHR (W-SHR) still achieved the highest AUC value.  As shown in Fig. \ref{fig:datasets}, The target vehicle distribution in HYDICE is very sparse while the target plane distribution in AVIRIS \uppercase\expandafter{\romannumeral2} is very close. It follows that the proposed W-SHR method is suitable for various distributions of targets in the entire image.

\section{CONCLUSION}
\label{sec:foot}

In this paper, a novel W-SHR method for target detection in HSIs is proposed. 
Firstly, a hierarchical background dictionary is constructed by concatenating global and local background dictionaries. 
And then, the issue of imbalanced dictionary quality is solved. 
The ROC curves and the AUC values on three real hyperspectral data demonstrate the superiority of the proposed W-SHR method. 
Based on the experimental results, the following conclusions can be drawn. 
1) Hierarchical background dictionary based on the global and local information can effectively improve the performance of sparse-based target detection methods.
2) Weighting scores from target and background dictionary according to the quality of dictionary can achieve good results in the case of unbalanced dictionary quality.


%
%
\footnotesize
\vspace{-0.3cm}
\setlength{\bibsep}{2.05ex}
\bibliographystyle{IEEEbib}
\bibliography{refs}

\begin{thebibliography}{10}

\bibitem{Du2019}
D.~Zhu, B.~Du, and L.~Zhang,
\newblock ``Target {Dictionary} {Construction}-{Based} {Sparse}
  {Representation} {Hyperspectral} {Target} {Detection} {Methods},''
\newblock {\em IEEE Journal of Selected Topics in Applied Earth Observations
  and Remote Sensing}, vol. 12, no. 4, pp. 1254--1264, 2019.

\bibitem{Du2019_Autoencoder}
S.~Chang, B.~Du, and L.~Zhang,
\newblock ``A {Sparse} {Autoencoder} {Based} {Hyperspectral} {Anomaly}
  {Detection} {Algorithm} {Using} {Residual} of {Reconstruction} {Error},''
\newblock in {\em Proc. IEEE International Geoscience and Remote Sensing
  Symposium}, 2019, pp. 5488--5491.

\bibitem{HSDD}
X.~{Lu}, W.~{Zhang}, and X.~{Li},
\newblock ``A {Hybrid} {Sparsity} and {Distance}-{Based} {Discrimination}
  {Detector} for {Hyperspectral} {Images},''
\newblock {\em IEEE Transactions on Geoscience and Remote Sensing}, vol. 56,
  no. 3, pp. 1704--1717, 2018.

\bibitem{CEM}
Q.~Du, H.~Ren, and C.~Chang,
\newblock ``A {Comparative} {Sstudy} for {Orthogonal} {Subspace} {Projection}
  and {Constrained} {Energy} {Minimization},''
\newblock {\em IEEE Transactions on Geoscience and Remote Sensing}, vol. 41,
  pp. 1525--1529, 2003.

\bibitem{MSD}
Louis~L Scharf and Benjamin Friedlander,
\newblock ``Matched {Subspace} {Detectors},''
\newblock {\em IEEE Transactions on Signal Processing}, vol. 42, no. 8, pp.
  2146--2157, 1994.

\bibitem{ACE}
X.~Jin, S.~Paswaters, and H.~Cline,
\newblock ``A {Comparative} {Study} of {Target} {Detection} {Algorithms} for
  {Hyperspectral} {Imagery},''
\newblock in {\em Proc. Algorithms and Technologies for Multispectral,
  Hyperspectral, and Ultraspectral Imagery XV}, 2009, vol. 7334, p. 73341W.

\bibitem{STD}
Yi. Chen, N.~M. asrabadi, and T.~D. Tran,
\newblock ``Sparse {Representation} for {Target} {Detection} in {Hyperspectral}
  {Imagery},''
\newblock {\em IEEE Journal of Selected Topics in Signal Processing}, vol. 5,
  no. 3, pp. 629--640, 2011.

\bibitem{SRBBH}
Y.~Zhang, B.~Du, and L.~Zhang,
\newblock ``A {Sparse} {Representation}-{Based} {Binary} {Hypothesis} {Model}
  for {Target} {Detection} in {Hyperspectral} {Images},''
\newblock {\em IEEE Transactions on Geoscience and Remote Sensing}, vol. 53,
  no. 3, pp. 1346--1354, 2014.

\bibitem{CSCR}
B.~Zhang W.~Li, Q.~Du,
\newblock ``Combined {Sparse} and {Collaborative} {Representation} for
  {Hyperspectral} {Target} {Detection},''
\newblock {\em Pattern Recognition}, vol. 48, no. 12, pp. 3904--3916, 2015.

\bibitem{HSSD}
B.~Du, Y.~Zhang, L.~Zhang, and D.~Tao,
\newblock ``Beyond the {Sparsity}-{Based} {Target} {Detector}: {A} {Hybrid}
  {Sparsity} and {Statistics}-{Based} {Detector} for {Hyperspectral}
  {Images},''
\newblock {\em IEEE Transactions on Image Processing}, vol. 25, no. 11, pp.
  5345--5357, 2016.

\bibitem{ODL}
J.~Mairal, F.~Bach, J.~Ponce, and G.~Sapiro,
\newblock ``Online {Dictionary} {Learning} for {Sparse} {Coding},''
\newblock in {\em Proc. International Conference on Machine Learning}, 2009,
  pp. 689--696.

\bibitem{BCCR_2019}
D.~Zhu, B.~Du, and L.~Zhang,
\newblock ``Binary-{Class} {Collaborative} {Representation} for {Target}
  {Detection} in {Hyperspectral} {Images},''
\newblock {\em IEEE Geoscience and Remote Sensing Letters}, 2019.

\end{thebibliography}

\end{document}